# Moiré superlattices of antimonene on a Bi(111) substrate with van Hove singularity and Rashba-type spin polarization

Tomonori Nakamura[1,2], Yitao Chen[1,3], Ryohei Nemoto[1], Wenxuan Qian[1,3], Yuto Fukushima[4], Kaishu Kawaguchi[4], Ryo Mori[4], Takeshi Kondo[4,5], Youhei Yamaji[1], Shunsuke Tsuda[6], Koichiro Yaji[6], and Takashi Uchihashi[1,3]*

[1] Research Center for Materials Nanoarchitectonics (MANA), National Institute for Materials Science, 1-1, Namiki, Tsukuba, Ibaraki 305-0044, Japan

[2] Okinawa Institute of Science and Technology Graduate University, 1919-1 Tancha, Onna-son, Kunigami-gun, Okinawa, 904-0495 Japan

[3] Graduate School of Science, Hokkaido University, Kita-10 Nishi-8, Kita-ku, Sapporo 060-0810, Japan

[4] Institute for Solid State Physics, The University of Tokyo, Kashiwa, Chiba 277-8581, Japan

[5] Trans-scale Quantum Science Institute, The University of Tokyo, Bunkyo-ku, Tokyo 113-0033, Japan

[6] Center for Basic Research on Materials (CBRM), National Institute for Materials Science, 3-13, Sakura, Ibaraki 305-0003, Japan

*Corresponding author
email: UCHIHASHI.Takashi@nims.go.jp

# Abstract

Moiré superlattices consisting of two-dimensional materials have attracted immense attention because of emergent phenomena such as flat band-induced Mott insulating states and unconventional superconductivity. However, the effects of spin-orbit coupling on these materials have not yet been fully explored. Here, we show that single- and double-bilayer antimony honeycomb lattices, referred to as antimonene, form moiré superlattices on a Bi(111) substrate due to lattice mismatch. Scanning tunnelling microscopy (STM) measurements reveal the presence of spectral peaks near the Fermi level, which are spatially modulated with the moiré period. Angle-resolved photoemission spectroscopy (ARPES) combined with density functional theory calculations clarify the surface band structure with saddle points near the Fermi level, which allows us to attribute the observed STM spectral peaks to the van Hove singularity. Moreover, spin-resolved ARPES measurements reveal that the observed surface states are Rashba-type spin-polarized. The present work has significant implications in that Fermi surface instability and symmetry breaking may emerge at low temperatures, where the spin degree of freedom and electron correlation also play important roles.

# Introduction

The advent of artificially stacked two-dimensional (2D) materials with moiré superlattices, which are induced by a lattice mismatch and/or a twist angle, has led to the development of new paradigms in condensed matter physics and materials science [1-3]. The successful fabrication of graphene/hexagonal boron nitride heterostructures and, more recently, twisted double layers of graphene have spawned a series of exciting discoveries, e.g., Hofstadter's butterfly and fractal quantum Hall effect [4-6], correlated insulating states and unconventional superconductivity [7-9], charge order [10], ferromagnetism and the quantum anomalous Hall effect [11-14]. The introduction of spin-orbit coupling (SOC) into moiré superlattices can lead to even richer emergent phenomena such as topological superconductivity, but these studies have been limited to transition metal dichalcogenides thus far [14-18]. The application of elemental 2D materials with strong SOCs may greatly expand the potential of moiré superlattices, but the exploration of such a possibility has been scarce thus far [19].

Bi and Sb are heavy VA elements with strong atomistic SOC and hence important ingredients in topological materials. The most stable form of bulk crystals is the A7 rhombohedral crystal structure, which consists of covalently bonded buckled honeycomb 2D layers (conventionally called bilayers) stacked by weak interlayer bonding [20-22]. This feature makes Bi and Sb atomic layers promising 2D materials beyond graphene [23-25]. They were theoretically predicted to become 2D topological insulators (quantum spin Hall insulators), and their topological edge states were found experimentally [26-32]. They are also reported to be resilient against air exposure and chemical processes and thus can potentially be used for various practical applications [23,24,33,34]. Atomic layers of Bi and Sb are commonly referred to as bismuthene [24,31] and antimonene [23,25,34-40], while this nomenclature is not completely correct because they are buckled and do not includes double bonds in their lattices [23]. Bismuthene and antimonene can be formed on an adequate substrate by molecular beam epitaxy (MBE) [31,34-37], in some cases leading to black phosphorus-like puckered layers [40,41].

Here, we show by scanning tunneling microscopy (STM) that single- and double-bilayer (BL) antimonene form moiré superlattices on a Bi(111) substrate due to lattice mismatch. They exhibit clear spectral peaks located near the Fermi level, which show distinctive behaviors regarding the moiré periodicity. While the peaks found for single-BL antimonene show only weak modulations in height, those for double-BL antimonene are split by ~100 mV and exhibit strong spatial modulations, suggesting localization due to the moiré superlattice. Angle-resolved photoemission spectroscopy (ARPES) and spin-resolved ARPES (SARPES) combined with density functional theory (DFT) calculations clarify the presence of saddle points near the Fermi level. This allows us to identify the origin of the STM spectral peaks as the van Hove singularity. These surface states are found to be Rashba-type spin polarized. The present work has significant implications for demonstrating that Fermi surface instability and symmetry breaking may emerge at low temperatures, where the spin degree of freedom and electron correlation are intimately involved.

## Results

### Atomic structures: STM measurements

All the experiments were carried out in ultrahigh vacuum (UHV) chambers with a base pressure of ~1×10$^{-10}$ mbar (see Methods). First, a Bi(111) thin film was grown on a clean Si(111) surface to 10 BLs by MBE [41,42.] For the ARPES/SARPES experiments with an imaging-type instrument (see below), Ge(111) substrates were used because of easiness of sample preparation. In this case, a 250 BL-thick Bi(111) film was grown on a clean Ge(111) surface [43,44]. These two recipes resulted in essentially the same Bi(111) surface structures and did not affect our conclusion. Here, we adopt the rhombohedral crystallographic notation to describe the plane index of the film [20,21]. The crystallinity of the prepared samples was confirmed with scanning tunneling microscopy (STM) and low-energy electron diffraction (LEED). The surface of the Bi(111) film was then covered by 1-2 BL Sb (for LEED patterns, see Supplementary Information A). Since Bi(111) and Sb(111) bilayers share the same buckled honeycomb structure with similar lattice constants of 0.454 nm and 0.431 nm, respectively, a Sb(111) film could grow epitaxially on a Bi(111) film [45,46]. However, the lattice constant of the free-standing Sb bilayer (1BL antimonene), which is predicted to be 0.408-0.412 nm [47,48], is significantly smaller than that of bulk Bi(111). This allows antimonene to grow nonepitaxially on Bi(111) and to form a moiré superlattice, which we indeed observe as follows.

The main panel of Fig. 1a shows a representative STM image of a Bi(111) surface covered with more than 1 BL of Sb. The lower terrace in the image (Region I) features a triangular lattice structure, which is a moiré superlattice made of 1BL antimonene (1BL Sb) on a Bi(111) surface. Although this superlattice includes defects and local deformations, the presence of a well-defined periodicity is clear from its fast Fourier transform (FFT) image (inset of Fig. 1a). From repeated experiments with different surface regions and samples, we determined the moiré lattice constant to be 4.70±0.30 nm (Supplementary Information B). On the upper terrace, there exists another region of the moiré superlattice with a longer periodicity (Region II). Since the height difference of ~0.4 nm between Regions I and II (Fig. 1b) is approximately equal to the height of the Sb bilayer (0.374 nm for bulk) [20], Region II is identified as 2BL antimonene (2BL Sb) on a Bi(111) surface. Its moiré lattice constant was determined to be 6.59±0.89 nm. The relatively large uncertainty is due to variations throughout different surface regions, presumably reflecting very small differences in energy. This 2BL Sb layer is bordered by another 1BL Sb layer (Region III), which is located in the upper-right corner of the image. Since they have almost the same topographic heights [20], the boundary (indicated by the dashed line) is identified as the location of a buried atomic step of the Bi(111) surface. Our repeated experiments indicate that antimonene layers grow from the step edges of Bi(111) surfaces. They also show that 2BL Sb layers begin to grow before 1BL Sb layers fully cover the whole surface, but 1BL Sb layers occupy ~70% of the total area when the nominal coverage $\theta$ of Sb is 1.0 BL. The same trend also applies to the growth of 2BL Sb. Therefore, 1BL and 2BL Sb are the dominant phases for $\theta$ = 1.0 BL and 2.0 BL, respectively (Supplementary Information C).

Magnified STM images of 1BL and 2BL Sb are displayed in Fig. 1c and Fig. 1d, respectively, where the Sb atomic lattices are clearly resolved. The moiré unit cells are indicated by the dashed parallelograms. These features are ascribed to the surface topography because they are reproduced with different bias voltages (for example, see Fig. 2a, d below).We determined the lattice constant of 1BL Sb to be 0.415±0.004 nm (Supplementary Information B). The fact that this value is greater than that of free antimonene (0.408-0.412 nm) [47,48] is attributed to the tensile strain exerted from the Bi(111) surface. Likewise, the lattice constant of 2BL Sb was determined to be 0.423±0.005 nm. This value is closer to that of bulk Sb(111) (0.431 nm) than that of 1BL Sb, suggesting lattice relaxation toward the bulk crystal. Combined with the moiré lattice constant determined above, the numbers of Bi and Sb atoms included in the moiré superlattice can be calculated. Assuming that the moiré unit cell consists of $N_{Bi} \times N_{Bi}$ Bi atoms and $N_{Sb} \times N_{Sb}$ Sb atoms per layer ($N_{Sb} = N_{Bi} + 1$), we find $N_{Bi}$ = 10, $N_{Sb}$ = 11 for 1BL Sb and $N_{Bi}$ = 13-17, $N_{Sb}$ = 14-18 for 2BL Sb. Our FFT analysis of STM images over an extended area reveals that there is no twisting between the moiré and Sb lattices on average (Supplementary Information B), although there are some local deviations due to deformations. Furthermore, Fig. 1c, d shows that the surface is divided into three characteristic regions in terms of topographic height. By comparing these observations to previous reports on related moiré superstructures [49-53], we can safely assign them to the regions of the AA, AB, and AC stacking sequences (Fig. 1e). In the AA stacking, all atoms in the two layers are vertically overlapped, while only half of them are in the AB and AC stacking layers. Because of the significant buckling of the honeycomb lattice, the vertical distance between the overlapping atoms in the AB stacking is greater than that in the AA stacking. This leads to the lowering of the top layer by an attractive force. Conversely, in the AC stacking, the vertical distance between the overlapped atoms is smaller than that in the AA stacking. This leads to the raising of the top layer by a repulsive force. As a result of structural relaxation, the areas corresponding to the AB and AC stacking regions expand and shrink, respectively [49,50]. These features are clearly observed in Fig. 1c, d.

**Electronic structures: STS measurements**

The electronic states of the moiré superlattices and their spatial modulations were investigated by scanning tunnelling spectroscopy (STS). First, for 1BL Sb on Bi(111), *dI/dV* spectra were taken at the center of the AA stacking region at five locations, and this process was repeated for the AB and AC stackings. The selected spectral sites are shown in the topographic STM image in Fig. 2a with red (AA), blue (AB), and green (AC) squares. Figure 2b shows the results of the STS measurements. The broken lines show individual *dI/dV* spectra, and the solid lines show the average values for the same stacking sequences, with their colors corresponding to those in Fig. 2a. The average of all the measured spectra are also shown by the solid black line. For all of these spectra, clear peak structures are noticeable near the zero bias voltage, with the full width at half maximum of $80 - 100$ mV. The spectral peaks at the AB sites are particularly conspicuous, while those at the AA and AC sites are relatively suppressed. More detailed information was obtained through line spectroscopy; *dI/dV* spectra were taken along a straight line connecting the centers of the AC, AB, AA, and AC sites in this sequence. The right panel of Fig. 2c shows a 2D plot of color-coded *dI/dV* spectra as a function of bias voltage and lateral distance from the starting point. In the left panel of Fig. 2c, the topographic profile along the line is also shown. We find that the spectral peak is fixed at approximately $0 - 30$ mV, while its intensity varies. This result strongly suggests the presence of delocalized states around the Fermi level that are weakly modulated by the moiré superlattice.

Figure 2d-f displays site-dependent *dI/dV* spectra for 2BL Sb on Bi(111) obtained in the same manner. The symbols and colors used in the figure follow the conventions of Fig. 2a-c. Figure 2e shows that spectral peaks are shifted from zero bias by 48 mV (AA site), -48 mV (AB site), and 144 mV (AC site) on average. The right panel of Fig. 2f shows that clear peak structures at approximately 40 mV and -100 mV are confined within the AA and AB regions, respectively. This result indicates the presence of multiple states near the Fermi level that are localized due to the moiré superlattice [54].

**Electronic structures : ARPES/ SARPES measurements**

To clarify the origin of the spectral peaks observed by STS, we performed laser-based high-resolution ARPES/SARPES measurements [55]. For simplicity, the data were analyzed based on the Brillouin zone of Bi(111) (Fig. 3a). First, we focused on the results for 1BL of Sb on a Bi(111) surface. Figure 3b shows a 2D plot of the ARPES intensity along the $\overline{\Gamma} - \overline{M}$ direction and as a binding energy $E_B$ (dark: high, bright: low). We can recognize two bands, denoted as $S_1$ and $S_2$, starting from an $E_B \cong 0.2$ eV and dispersing upward. These bands disappear at approximately $k_x = 0.05$–$0.1$ Å$^{-1}$ by crossing the Fermi level but seem to disperse downward and reappear at approximately $k_x = 0.4$–$0.5$ Å$^{-1}$. The band dispersions determined from the plot are highlighted with red dashed curves. The signals are better visualized by the SARPES signal plotted for the range of -0.22 Å$^{-1} < k_x < +0.22$ Å$^{-1}$, where the intensity and the spin polarization in the y direction are indicated by brightness (dark: high, bright: low) and color (red: positive, blue: negative), respectively. The maximum spin polarization of the photoelectron is ~0.6. The spin polarizations of the $S_1$ and $S_2$ bands are opposite to each other and are antisymmetric with respect to $k_x = 0$ Å$^{-1}$. This is characteristic of Rashba-type spin-polarization, which will be discussed later. Figure 3c shows a similar 2D plot of the ARPES intensity along the $\bar{\Gamma} - \bar{K}$ ($k_y$) direction. The $S_1$ and $S_2$ bands are also noticeable (highlighted with red dashed lines), but the $S_2$ band reaches a local maximum near the Fermi level at approximately $k_y = 0.1$- $0.15$ Å$^{-1}$ and then disperses downward. For better visibility, the same ARPES data in Fig. 3b, c are displayed in Fig. D2a, b without red dashed lines (Supplementary Information D).

Figure 3e shows the 2D plot of the ARPES intensity measured near the Femi level ($E_B = 0.02$ eV) in the $k_x$ - $k_y$ space, which gives the Fermi surface contour. The central ring and a surrounding star-like structure are clearly noticeable and can be identified as the $S_1$ and $S_2$ bands, respectively. Notably, some parts of the $S_2$ band appear very weak due to the anisotropic transfer during the photoemission process. By referring to the band dispersions in Fig. 3b-d, we can identify the areas indicated by the red ellipses as saddle points, where the $S_2$ band takes a local maximum in the $\bar{\Gamma} - \bar{K}$ direction ($k = 0.1$- $0.15$ Å$^{-1}$, $E_B \sim 0$ eV; see Fig. 3c) and a local minimum in the orthogonal direction. This means that the van Hove singularity exists at the Fermi level [56] and explains the origin of the zero bias peak for 1BL Sb/Bi(111) described above. To confirm this result, we also performed Fermi surface mapping with an imaging-type ARPES instrument, which allows us to access a larger momentum space at a faster speed (Fig. 3f) [44]. The acquired Femi surface well reproduces the features observed in Fig. 3e while better reflecting the sixfold symmetry expected from the $C_3$ and time-reversal symmetries of the present system. We note that the band structure and the Fermi surface resemble those of Bi(111) and Sb(111) surfaces [21,42,43,57,58], while saddle points are absent near the Fermi level in the latter cases.

Because of the space-inversion symmetry breaking at surface and strong SOC of Sb and Bi, Rashba-type spin polarization is expected. To confirm this, the distribution of spin polarization in momentum space was investigated with the same imaging-type instrument for 1BL Sb/Bi(111). Figure 3g shows a 2D plot of the SARPES intensity and spin polarization in the y direction ($P_y$) measured near the Fermi level ($E_B = 0.03$ eV). The observed signal is mostly attributed to the $S_2$ band, the location of which is reproduced from Fig. 3f (red solid lines). Along the $k_x$ axis (white dashed line) and near the $\bar{\Gamma} - \bar{K}$ lines at ±60° to the $k_x$ axis (black dashed lines), $P_y$ is reversed when momentum is reversed with respect to the $\bar{\Gamma}$ point (see also Supplementary Information D, Fig. D3), although their intensities are different. The variations in the intensity can be attributed to the fact that SARPES (more generally, ARPES) signal is strongly dependent on the transition matrix in the photoemission process. These results are consistent with Rashba-type spin polarization. We should note that the actual distribution of spin polarization deviates from the ideal vortical form, as indicated by the reversal of $P_y$ with respect to the black dashed lines. An analogous behavior was also predicted and observed for a clean Bi(111) surface with a giant Rashba splitting [44,59,60].

We also performed ARPES/SARPES measurements of 2BL Sb on a Bi(111) surface (Supplementary Information D, Fig. D1). These results are nearly identical to those for 1BL Sb/Bi(111) (Fig. 3), but the observed ARPES signals are clearer than those for 1BL Sb/Bi(111). This difference may be attributed to the better moiré periodicity observed with STM (Fig. 1a).

## Electronic structures : DFT calculations

Ab initio calculations of the electronic structure of Sb/Bi(111) moiré superlattices are difficult because of the large number of heavy atoms involved within a moiré unit cell. To circumvent this problem, we carried out DFT calculations based on an epitaxial model consisting of 1BL Sb(111) on 5BL Bi(111) (Methods). Although this model does not include the effect of moiré periodicity, it can account for the overall band structures within the Bi(111) Brillouin zone. The structural relaxation within each stacking region in the actual moiré structure (Fig. 1c, d) rationalizes this treatment. Figure 4b shows the band diagram calculated for the epitaxial model with the AB stacking (Fig. 4a). The orange and blue rectangles correspond to the same marked areas in Fig. 3b,c. The sizes of the purple (light blue) circles represent the contributions of the top Sb (Bi) BL. Overall, the two bands starting from the $\bar{\Gamma}$ point below the Fermi level (designated by the red dashed lines) are mostly derived from the top Sb BL, indicating that they can be preferentially detected in surface-sensitive STM and ARPES measurements. Judging from their dispersions, they can be assigned to the $S_1$ and $S_2$ bands identified above (Fig. 3b-d). Other bands with negligible contribution of Sb are assigned to the Bi(111) surface states on the back side of the model and thus are irrelevant to our experimental data. The same calculations for the AA and AC stackings give very similar band structures near the $\bar{\Gamma}$ point and around the Fermi level (Supplementary Information E, Fig. E2b and Fig. E3b). Therefore, the presence of saddle points near the Fermi level is theoretically confirmed. The energies $E$-$E_F$ and momenta ($k_x$, $k_y$) of the saddle points are summarized in Table E1 of Supplementary Information E. These features are reflected in the projected density of states (PDOS) on the top Sb BL (Fig. 4g). The three sharp peaks indicated by the arrows at $E - E_F = 0.02 - 0.05$ eV, corresponding to the AA, AB and AC stackings, are due to the van Hove singularity of the saddle point. The peak energies are very close to one another, reproducing the STS results shown in Fig. 2b,c. Assuming that the Fermi level is aligned near these peaks in real samples, we plot the Fermi surface contour of 1BL Sb(111)/5BL Bi(111) with AB stacking at $E - E_F = -0.02$ eV (Fig. 4c). The central rings and a surrounding star-like structure are consistent with the ARPES results (Fig. 3e, f). The spin structures at the Fermi surface calculated for this model is consistent with the SARPES plot in Fig. 3g (Supplementary Information E, Fig. E4).

The same calculations were also conducted based on an epitaxial model of 2BL Sb(111) on 5BL Bi(111) (Fig. 4d). The band structures and the Fermi surface ($E - E_F$ = -0.02 eV) obtained for the AB stack (Fig. 4e, f) resemble those obtained for the 1BL Sb(111) model (Fig. 4b, c) as well as the ARPES results (Supplementary Information D, Fig. D1). This is also the case for the AA and AC stackings (Supplementary Information E, Fig. E2d-f and Fig. E3d-f). However, the PDOSs around the Fermi level calculated for the AA, AB, and AC stacks exhibit more separated energies (Fig. 4h). Qualitatively, these results are in line with the STS data (Fig. 2e, f), but there are some clear discrepancies; e.g., the sharp peak at $E - E_F$ = 0 eV for the AC stacking has no corresponding structure in the STS data (Fig. 2e). This difference may be attributed to incomplete structural optimization of the 2BL Sb model, which results from our simplified models of fixing the locations of the Bi atom to those of the bulk Bi crystal (see Methods).

## Discussion

We calculated the band structures of antimonene on a Bi(111) surface based on epitaxial models of 1BL Sb(111)/5BL Bi(111) and 2BL Sb(111)/5BL Bi(111), which successfully explained our STM and ARPES data. Obviously, the adopted models are rather crude and do not include the effects of moiré modulations. The fact that all stacking sequences AA, AB, and AC result in qualitatively identical electronic band structures may also explain the success of the present models. Experimentally, the 1BL-Sb/Bi(111) moiré superlattice has nearly no spectral modulation (Fig. 2c), while the 2BL-Sb/Bi(111) moiré superlattice has a modulation of $0.1 - 0.2$ eV (Fig. 2f). Considering that the transfer integrals between the atomic orbitals in neighboring layers of Bi and Sb are $0.3 - 1.4$ eV [20], the influences of moiré superlattices are unexpectedly small. It suggests that it is not simply determined by the interlayer coupling strength.

Since Sb and Bi are isovalent, replacing Bi by Sb at the top layer simply leads to variations in electrostatic potential. It is thus possible to interpret the present $S_2$ band, featuring the saddle points near the Fermi level, as a result of continuous deformation of the $S_2$ band of a clean Bi(111) surface [61]. We also note that Bi is a topologically nontrivial semimetal and that its (111) surface states are protected [43,62]. Based on the same ground, the surface states of antimonene on Bi(111) can be topologically equivalent to those on Bi(111). In this case, the formation of an energy gap within the surface states is prohibited because it would violate the requirement that the bulk conduction and valence bands be continuously connected by surface states [63,64]. Indeed, our DFT calculations based on a moiré superlattice model indicate that there are no moiré-induced energy gaps at the zone boundary (see Supplementary Information F, Fig. F1). The result may be attributed to the topological protection discussed above. In this case, (high-order) van Hove singularities are predicted to emerge at the $\overline{K}$ point of the moiré Brillouin zone [63], which calls for a future study.

Here we briefly mention the possibility of alloying of the top layer. It is widely known that Sb and Bi can form alloys with arbitrary ratios, and thus Sb and Bi atoms may be mixed to some extent. However, since the alloy formation does not change the A7 rhombohedral crystal structure of bulk [65], it is highly likely to retain the buckled honeycomb lattice of antimonene. Since Sb and Bi are isovalent, the main effect of the mixing should be a simple potential modulation, which may slightly shift the band energy. Nevertheless, considering the good agreement between the band structure obtained by ARPES and that of our theoretical model, the alloying effect can be neglected for the current data set.

Finally, we discuss the implications of the findings in the present work. Generally, the presence of the van Hove singularity means a logarithmic divergence of the density of states and an enhancement of Coulomb interactions. Tuning the Fermi level to a van Hove singularity point can lead to a variety of symmetry-broken phases at low temperatures, such as unconventional superconductivity, charge density waves, ferromagnetism, charge order, and nematicity; these phases have been discussed within the context of high-$T_c$ cuprates, graphene, kagome metals, etc. [56,66-69]. Since the van Hove singularities in our systems are located close to the Fermi level, their tuning must be technically viable through gate voltage or molecular doping [70]. Among the possible low-temperature phases, superconductivity is the most likely because of the presence of Rashba-type spin polarization and the resulting spin-momentum locking [63]. In this case, the absence of space inversion symmetry should lead to a spin singlet-triplet mixed state [71]. We note that electron–phonon coupling, which is responsible for superconductivity, is likely to be enhanced here, as discussed for granular Bi films [72,73]. Regarding 2BL Sb/Bi(111), moiré-induced electron confinement and enhanced electron correlation can further enrich the low-temperature physics beyond the simple van Hove singularity scenario. The inclusion of Rashba-type spin polarization in the moiré system is an open and intriguing problem. Experimentally, preparation of a high-quality sample with less structural defects and a better moiré periodicity would be essential, since it should strengthen the divergent behavior of the van Hove singularity. Thus, the present Sb/Bi(111) moiré superlattices will offer a new playground for investigating the role of the spin degree of freedom in van der Waals materials.

## Methods

### Sample preparation

The Si(111) substrates were cleaned by direct current heating at 1250 °C for 10 seconds. After repeating the cycle several times, 7×7 clean surfaces were obtained. The Ge(111) substrates were cleaned by repeated Ar⁺ sputtering and annealing several times to obtain 2×1 surfaces. Both surfaces were confirmed by observing sharp LEED spots. Bi(111) films were then grown by MBE on Si(111)-7×7 surfaces to 10 BLs or on Ge(111)-2×1 surfaces to 250 BLs at room temperature. To improve the flatness of the film, the Bi films were annealed at approximately 190 °C for 5 min. Subsequently, Sb was deposited on the Bi(111) surfaces at room temperature to form moiré antimonene. The crystallinity of the sample was improved by mild annealing at approximately 100 °C.

### STM measurements

The STM measurements were conducted at 78 K and 4.6 K with a Nanonis controller Mimea BP5e. Topographic images were obtained in constant current mode. The lateral scale of the STM images was calibrated through observation of Bi(111) surfaces by assuming that the lattice constant is equivalent to that of a bulk crystal (0.454 nm). STS measurements were conducted with a built-in lock-in amplifier with a typical bias voltage modulation of 20 mV at 477 Hz.

### ARPES measurements

ARPES and SARPES measurements were performed at National Institute for Materials Science (NIMS) and at the Institute for Solid State Physics (ISSP), University of Tokyo. For the NIMS measurements, we employed a momentum microscope equipped with an imaging spin detector [44,74]. A 10.9-eV laser was used as the excitation light. For the measurements at the ISSP, the photoelectrons excited by a 6.994-eV laser were analyzed by a hemispherical photoelectron analyzer equipped with an ultralow-speed electron diffraction spin detector [55]. For both measurements, samples were prepared *in situ*. The sample temperature during the measurements was 30 K.

### DFT calculations

To calculate the electronic band structures, we adopted epitaxial models of 1BL Sb(111)/5BL Bi(111) and 2BL Sb(111)/5BL Bi(111). The locations of the Bi atoms and those of the Sb atoms in the in-plane directions were fixed to those of the bulk Bi crystal. The out-of-plane Bi atom positions follow the structure for the Bi(111) thin film used in the literature [61]. Here, the Bi atom locations in the top Bi BL (adjacent to the Sb BLs) are modified, while the other Bi atoms follow the bulk Bi structure. In contrast, the locations of the Sb atoms in the out-of-plane direction were set equal to those at the centers of the AA, AB and AC stacking regions of the numerically optimized moiré superlattices (indicated by the red, blue and green circles, respectively, in Fig. E1; see Supplementary Information E). To optimize the 1BL Sb(111)/5BL Bi(111), we prepared a single 11×11 supercell of the Sb(111) BL on five vertically stacked 10×10 supercells of the Bi(111) BL. While the locations of the Bi atoms are fixed as in the Bi(111) thin film [61], the locations of the Sb atoms are optimized by utilizing a neural network potential, PFP (without U) version 5.0.0, on Matlantis (https://matlantis.com/) [75]. Similarly, to optimize the 2BL Sb(111)/5BL Bi(111), we prepared two 14×14 supercells of the Sb BL on five 13×13 supercells of the Bi BL and optimized the positions of the Sb atoms.

The noncollinear DFT calculations for the epitaxial models are performed by OpenMX version 3.9 [76-79] with the PBE exchange correlation functional and spin–orbit coupling. We use the 15×15×1 $k$-point grid in the first Brillouin zone for self-consistent field (SCF) calculations. After the SCF calculation, the density of states is obtained on a 256×256×1 $k$-point grid by the tetrahedron method. The energy cutoff is set to 100 or 200 Ry. The dependence of the band structure and density of states on the $k$-point grid and energy cutoff was examined to ensure convergence.

## Data availability

The datasets generated during and/or analyzed during the current study are available from the corresponding author upon reasonable request.

## Acknowledgments

The authors thank S. Yoshizawa, F. Arai, S. Takezawa, H. Tanaka, A. Harasawa, and T. Iimori for their technical support during the STM and ARPES experiments. This work was supported financially by JSPS KAKENHI (Grant Numbers 20H05621, 22H01961, 20K15133, 22H01183, 23H03818, 23H04524), the World Premier International Research Center (WPI) Initiative on Materials Nanoarchitectonics, MEXT, Japan and the Innovative Science and Technology Initiative for Security Grant Number JPJ004596, ATLA, Japan.

## Author contributions

T.N. and T.U. conceived the experiment, and T.N., T.U., K.Y. and Y.Y. wrote the manuscript. T.N., R.N. and W.Q. carried out the STM measurements, and T.N. analyzed the data. T.N., K.Y., Y.F., K.K., and R.M. carried out (S)ARPES measurements at ISSS under the supervision of T.K. Y.C., K.Y., and S.T. carried out (S)ARPES measurements at NIMS. K.Y. and S.T. analyzed the ARPES/SARPES data. Y.Y. performed the DFT calculations. All the authors discussed the results and contributed to finalizing the manuscript.

## Competing interests

The authors declare no competing interests.

## Additional information

### Supplementary information

The supplementary materials A-F are available online. Correspondence and requests for materials should be addressed to T. Uchihashi.

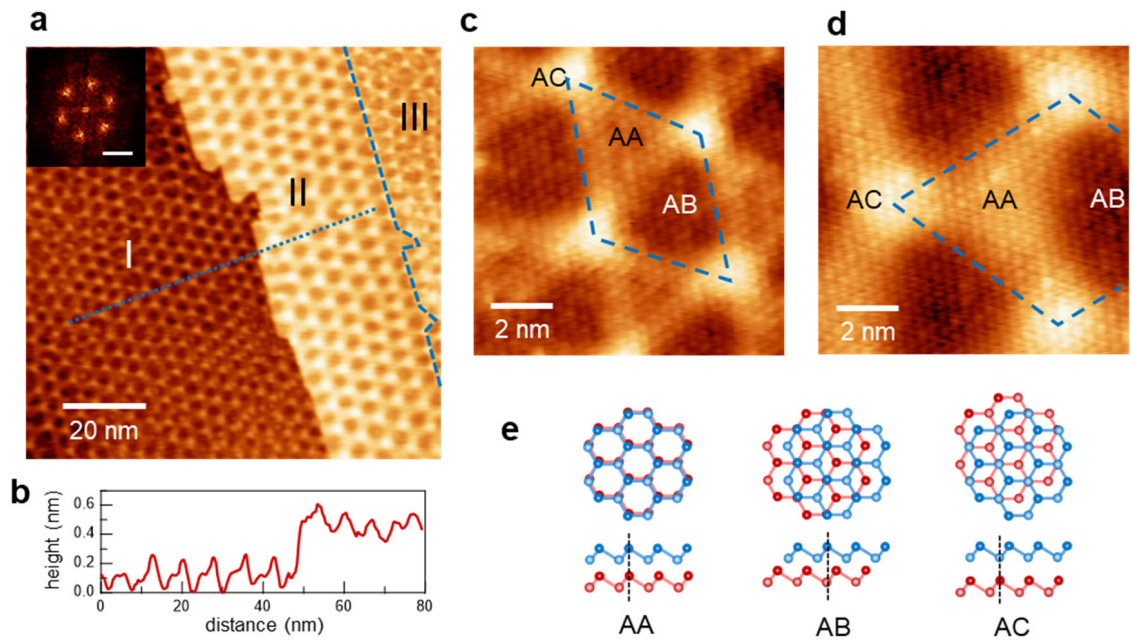

**Fig. 1 | Topographies and atomic structures of antimonene/Bi(111) moiré superlattices.**
**a** STM topographic image of 1BL and 2BL Sb grown on a Bi(111) film (Sample bias voltage: $V_s$ = -1.0 V, Tunnelling current: $I_t$ = 100 pA). Regions I and III correspond to 1BL Sb while Region II to 2BL Sb. The inset shows a FFT transform of a surface area belonging to Region I (scale bar: 0.3 nm⁻¹). **b** Height profile along the dotted line in **a**. **c, d** Atomic resolution images of 1BL Sb (**c**) and 2BL Sb (**d**) on a Bi(111) surface. ($V_s$ = -50 mV, $I_t$ = 500 pA (**c**), $V_s$ = -50 mV, $I_t$ = 300 pA (**d**)). The dashed parallelograms indicate the unit cells of the moiré superlattices. **e** The top and side views of the AA, AB and AC stacking sequences (blue spheres: Sb, red spheres: Bi). The vertical dashed lines shows the alignment of the Sb and Bi atoms.

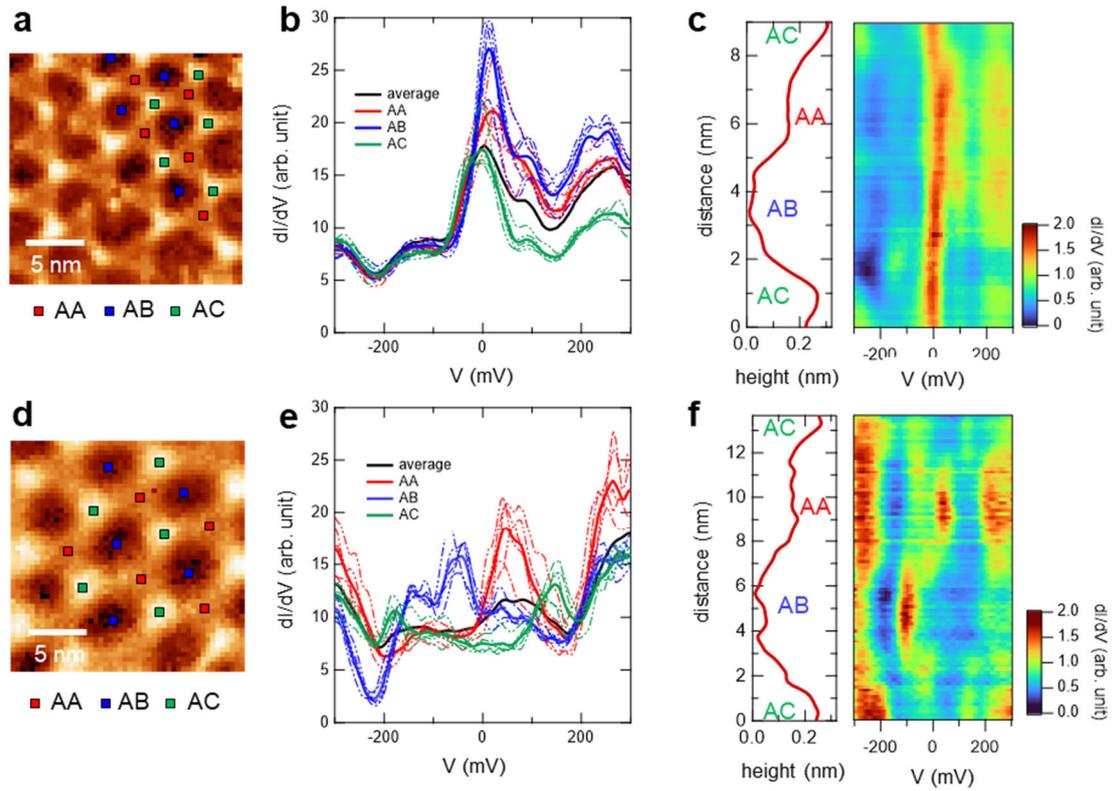

**Fig. 2 | Local electronic structures of antimonene/Bi(111) moiré superlattices near the Fermi level. a** STM topographic image of 1BL Sb/Bi(111) moiré superlattices ($V_s$ = -300 mV, $I_t$ = 100 pA). The red (AA), blue (AB), and green (AC) squares show the locations for the STS measurements. **b** $dI/dV$ spectra obtained at individual locations shown in **a** (broken lines) and their averages (solid lines) taken for the same stacking sequences. The red, blue, and green lines correspond to AA, AB, and AC stackings, respectively. The black solid line is the average of all the spectra. **c** (Right panel) 2D plot of color-coded $dI/dV$ spectra of 1BL Sb/Bi(111) moiré superlattices, which was taken along a straight line connecting the centers of AC, AB, AA, and AC regions ($V_s$ = -300 mV, $I_t$ = 100 pA). The vertical axis represents the lateral distance from the starting point. (Left panel) Topographic cross section along the measurement line. **d, e, f** Results for 2BL Sb/Bi(111) moiré superlattices obtained in the same manner with the identical imaging parameters as in **a, b, c**.

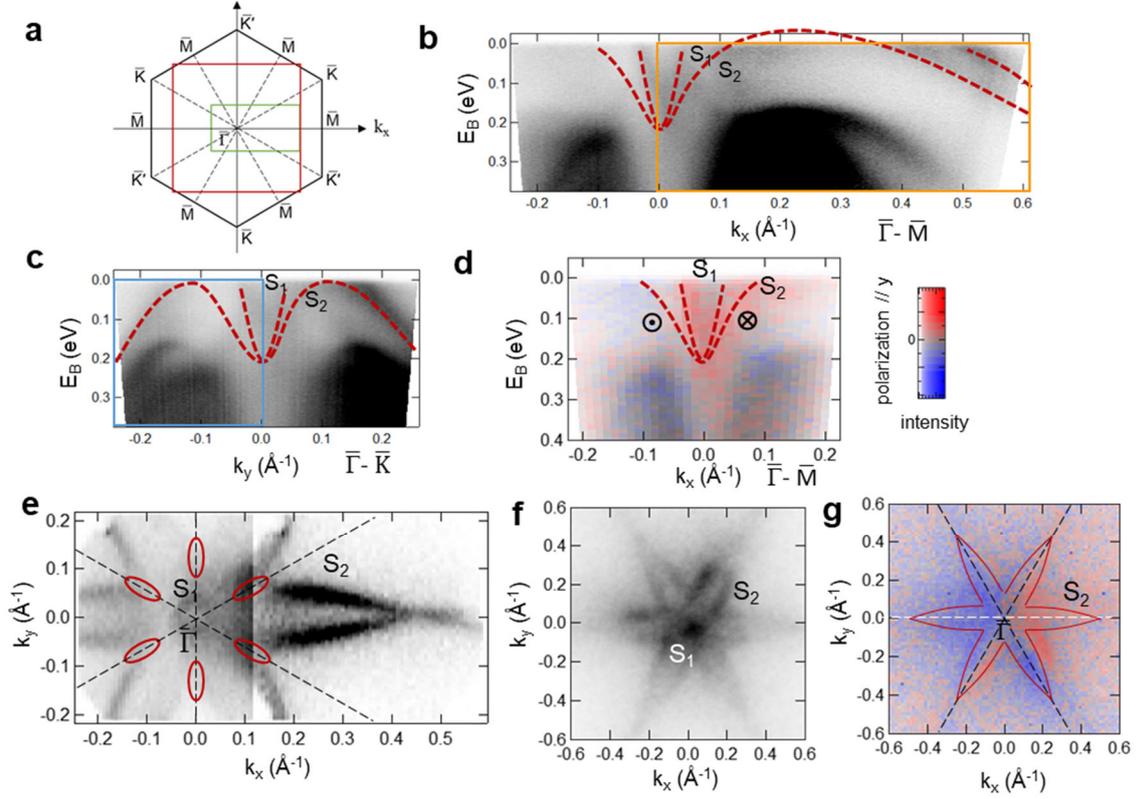

**Fig. 3 | Electronic and spin structures of 1BL antimonene/Bi(111) moiré superlattices in the momentum space. a** Bi(111) Brillouin zone and the high symmetry points $\bar{\Gamma}$, $\bar{M}$, $\bar{K}$, $\bar{K}'$. The green rectangle and the red square correspond to the areas for **e** and **f, g**, respectively. **b** 2D plot of ARPES intensity as a function of momentum $k_x$ along the $\bar{\Gamma} - \bar{M}$ direction and as a binding energy $E_B$. **c** The same plot as in **b** along the $\bar{\Gamma} - \bar{K}$ direction. The orange and blue rectangles correspond to those shown in Fig. 4a. **d** 2D plot of the SARPES signal as a function of momentum $k_x$ along the $\bar{\Gamma} - \bar{M}$ direction and as a binding energy $E_B$. The intensity and the spin polarization in the y direction are indicated by brightness (dark: high, bright: low) and color (red: positive, blue: negative), respectively. In **b-d**, the dispersions of $S_1$ and $S_2$ bands are indicated by the red dashed lines. **e** 2D plot of ARPES intensity measured near the Femi level ($E_B = 0.02$ eV) in the $k_x$ - $k_y$ space, which gives the Fermi surface contour. The plot includes two data sets obtained in separate runs, which causes an apparent discontinuity at $k_x = 0.12$ Å$^{-1}$. The dashed lines indicate the $\bar{\Gamma} - \bar{K}$ directions. The red ellipsoids show the locations of the saddle points of the $S_2$ band. **f** The same plot as in **e** in a larger momentum space, which was obtained with the imaging-type instrument. **g** 2D plot of SARPES intensity and spin polarization in the y direction measured near the Fermi level ($E_B = 0.03$ eV) obtained with the imaging-type instrument. The intensity and the spin polarization in the y direction are plotted as in **d**. The Fermi surface contour of the $S_2$ band determined from **f** is shown with red solid lines. The black and white dashed lines indicate the $\bar{\Gamma} - \bar{M}$ directions.

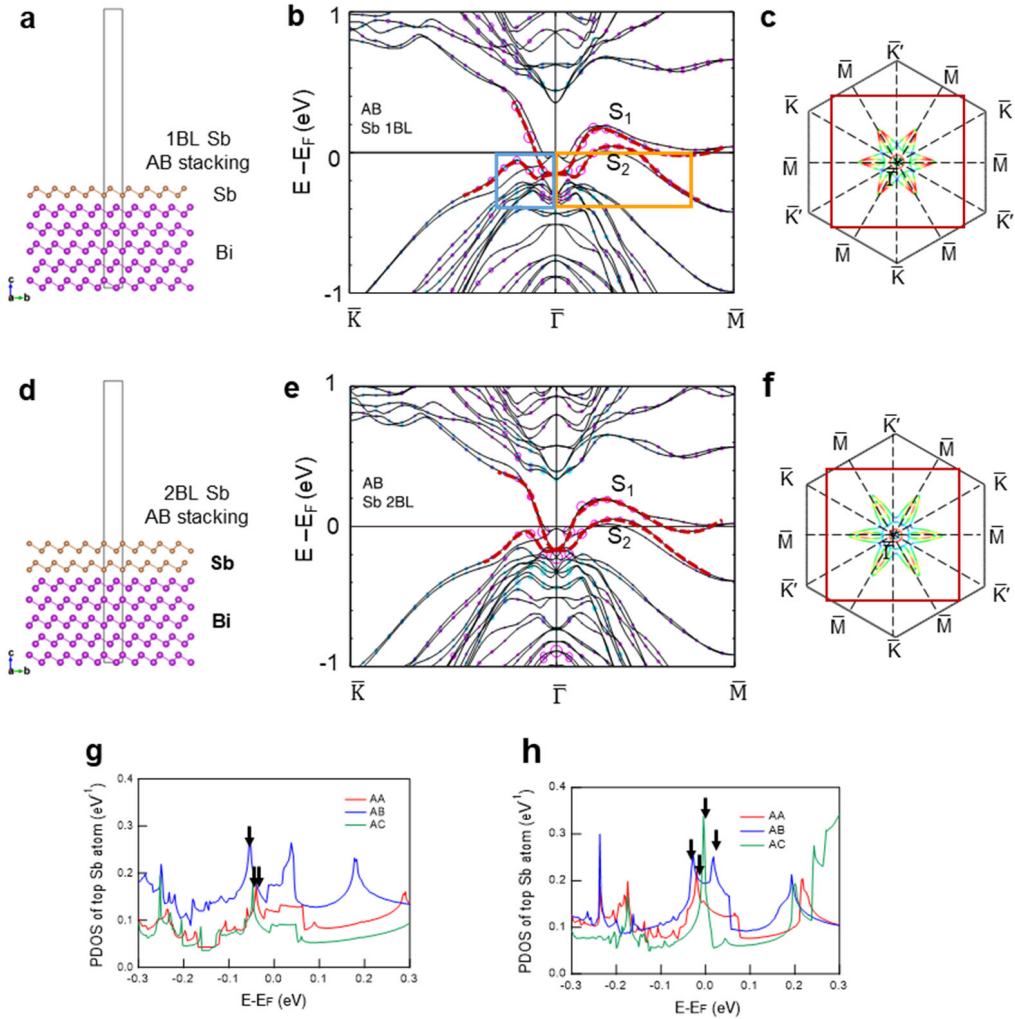

**Fig. 4 | DFT band structure calculations for the Sb(111)/Bi(111) epitaxial models. a, d** Atomic structure models for 1BL Sb(111)/5BL Bi(111) (**a**) and 2BL Sb(111)/5BL Bi(111) (**d**) with the AB stacking. The solid rectangles are the unit cells for calculations. **b, e** Band dispersions along the $\bar{K} - \bar{\Gamma} - \bar{M}$ direction calculated with the 1BL Sb(111)/5BL Bi(111) model (**b**) and the 2BL Sb(111)/5BL Bi(111) model (**e**) with AB stacking. The orange and blue rectangles correspond to those shown in Figs. 3b and 3c. The sizes of the purple (light blue) circles represent the contributions of the top Sb (Bi) BL. **c, f** Fermi surfaces ($E - E_F$ = -0.02 eV) calculated for the 1BL Sb(111)/5BL Bi(111) model (**c**) and the 2BL Sb(111)/5BL Bi(111) model with AB stacking (**f**). The colors represent the Fermi velocity (red: fast, blue: slow). The red squares correspond to the regions displayed in Fig. 3f, g. **g, h** Projected density of states (PDOS) on the top Sb BL calculated for the 1BL Sb(111)/5BL Bi(111) model (**g**) and the 2BL Sb(111)/5BL Bi(111) model (**h**).

# Supplementary Information for "Moiré superlattices of antimonene on a Bi(111) substrate with van Hove singularity and Rashba-type spin polarization"

Tomonori Nakamura[1,2], Yitao Chen[3], Ryohei Nemoto[1], Wenxuan Qian[1,3], Yuto Fukushima[4], Kaishu Kawaguchi[4], Ryo Mori[4], Takeshi Kondo[4,5], Youhei Yamaji[1], Shunsuke Tsuda[6], Koichiro Yaji[6], and Takashi Uchihashi[1,3]

[1] Research Center for Materials Nanoarchitectonics (MANA), National Institute for Materials Science, 1-1, Namiki, Tsukuba, Ibaraki 305-0044, Japan

[2] Okinawa Institute of Science and Technology Graduate University, 1919-1 Tancha, Onna-son, Kunigami-gun, Okinawa, 904-0495 Japan

[3] Graduate School of Science, Hokkaido University, Kita-10 Nishi-8, Kita-ku, Sapporo 060-0810, Japan

[4] Institute for Solid State Physics, The University of Tokyo, Kashiwa, Chiba 277-8581, Japan

[5] Trans-scale Quantum Science Institute, The University of Tokyo, Bunkyo-ku, Tokyo 113-0033, Japan

[6] Center for Basic Research on Materials (CBRM), National Institute for Materials Science, 3-13, Sakura, Ibaraki 305-0003, Japan

## Contents

## A. LEED patterns of antimonene on Bi(111) surfaces

LEED measurements were conducted for Bi(111) surfaces before and after deposition of Sb at room temperature (before: Fig. A1a, after: Fig. A1b-f). First, the deposition of Sb induces fine features around the principal LEED spots of Bi(111) (Fig. A1b-d), which are attributed to the moiré superlattices of 1BL antimonene. Increase in deposition time results in formation of broader spots (Figure A1e,f). The deposition time was converted to the Sb coverage through STM imaging of the prepared samples. Figure A1c (30 min deposition) corresponds to the coverage of 1.0 BL Sb.

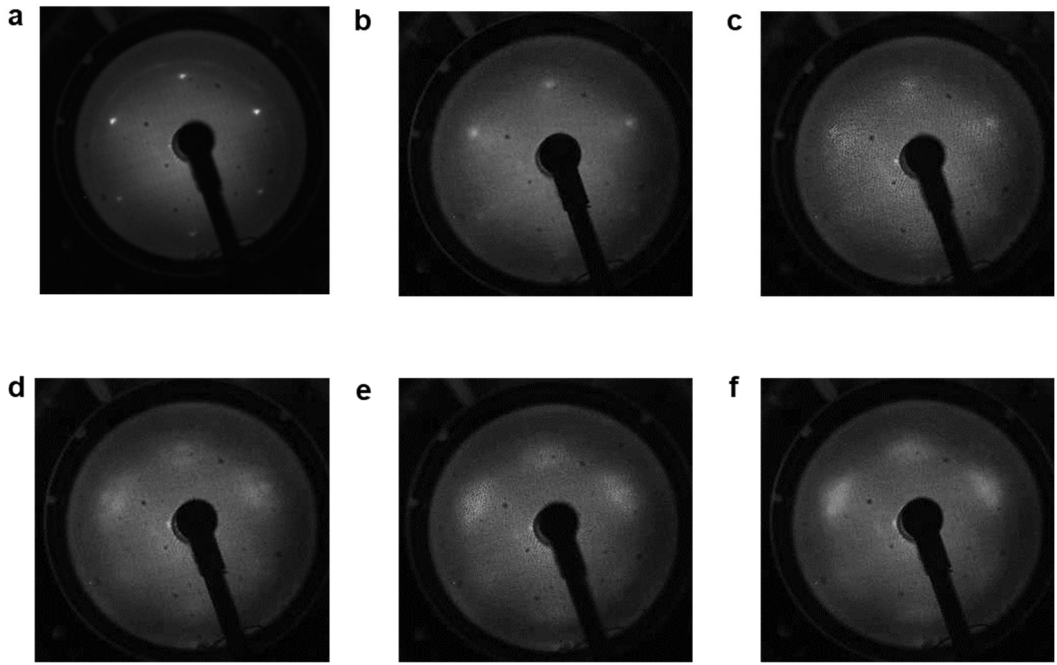

**Fig. A1 | LEED patterns of Sb on Bi(111) surfaces. a** Bi(111) clean surface. **b-f** Bi(111) surfaces after deposition of Sb for 20 min (0.67 BL Sb) (**b**), 30 min (1.0 BL Sb) (**c**), 40 min (1.33 BL Sb) (**d**), 50 min (1.67 BL Sb) (**e**), 60 min (2.0 BL Sb) (**f**). The beam energy was set at 46.5 eV.

## B. Determination of the atomic structures of antimonene on Bi(111) and the moiré superlattices

To determine the lattice constant of Sb layers on Bi(111) and the moiré periodicities, fast Fourier transform (FFT) images (Fig. B1b, c, e ,f) are taken from topographic image of 1BL Sb on Bi(111) (Fig. B1a) and 2BL Sb on Bi(111) (Fig. B1d). Two kinds of spots with the hexagonal symmetry are shown in FFT images. These spots are consistent with the Sb(111) surface and the moiré superlattice in topographic image. The outer spots indicated by yellow circles (Fig. B1b,e) correspond to the atomic lattice of Sb, while the inner spots indicated by red arrows (Fig. B1c, f) to the moiré superlattice. They are aligned in the same directions, revealing that there is no twisting between the Sb and Bi atomic layers. The lattice constant of the Sb lattice and the moiré period were calculated from the distance between the spots in FFT image. Our repeated experiments lead to a lattice constant of 0.415±0.004 (0.423±0.005) nm for the 1BL (2BL) Sb lattice. The moiré periodicity is 4.70±0.30 (6.59±0.89) nm for 1BL (2BL) Sb. The STM images were calibrated by assuming that the lattice constant of a 10BL Bi(111) clean surface is equal to that of a bulk crystal (0.454 nm).

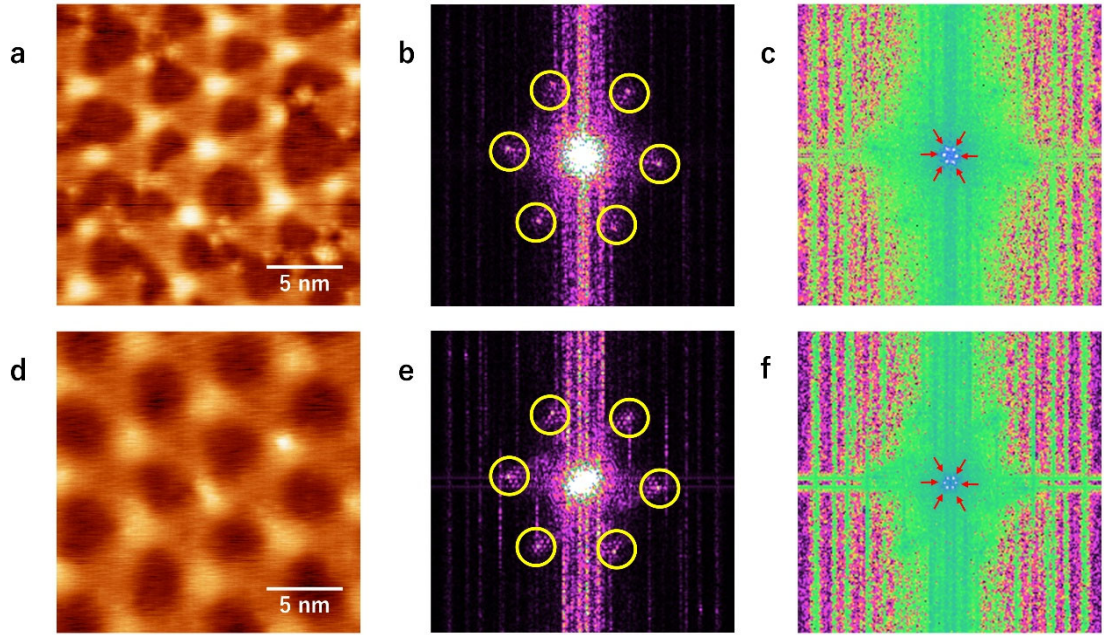

**Fig. B1 | STM images of antimonene on Bi(111) and their FFT analysis. a, d** STM topographic images of 1BL (**a**) and 2BL Sb (**d**) grown on a Bi(111) film (Sample bias voltage: $V_s$ = 10 mV, Tunnelling current: $I_t$ = 300 pA). **b, e** Fast Fourier transform images of **a** and **d**. The spots indicated by the yellow circles correspond to the atomic lattice of Sb. **c, f** Fast Fourier transform images of **a** and **d** displayed with a different colour scale. The spots indicated by the red arrows correspond to the moiré superstructures.

## C. Growth behaviour of antimonene

A Bi(111) film was grown by MBE on Si(111)-7×7 surfaces to 10 BL at the room temperature. To improve the flatness of the film, the Bi films was annealed around 190 °C for 5 min. A STM image shows flat Bi(111) terraces with step edge running along the three-fold crystallographic orientations (Fig. C1a). Subsequently, Sb was deposited on Bi(111) surfaces at the room temperature to form moiré antimonene. The first layer of antimonene was found to grow from step edges of Bi(111) surfaces toward to the lower side of the step, while the second layer of antimonene from those of the first layer (Fig. C1b, c). These observations indicate a step-flow growth mechanism of antimonene.

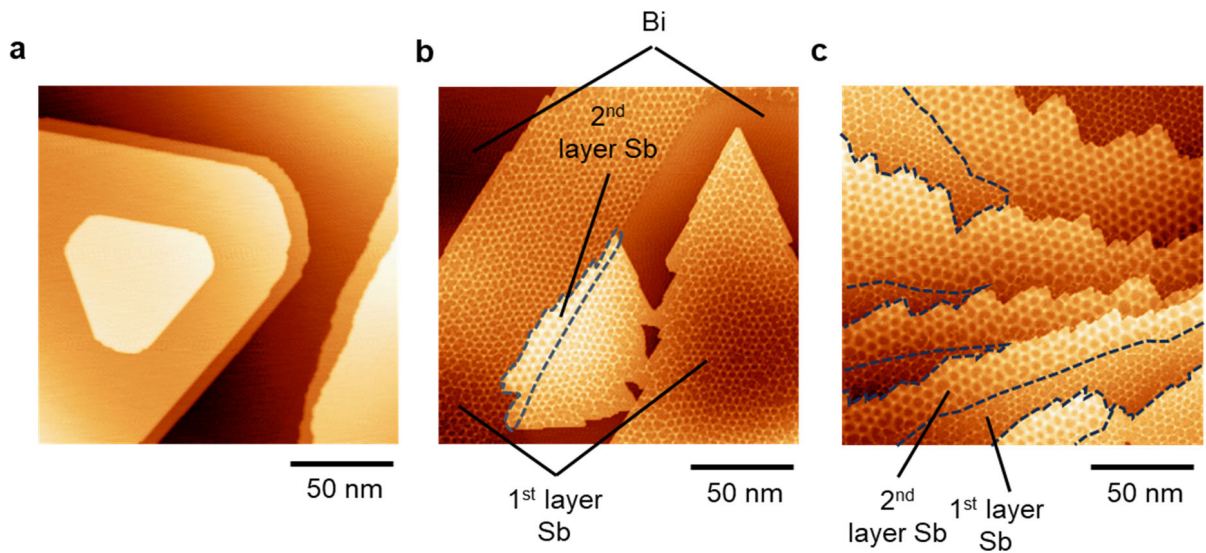

**Fig. C1 | Growth behaviour of antimonene on a Bi(111) surface. a** Topographic STM images of 10 BL Bi(111) on Si(111). **b, c** Topographic STM images of 10 BL Bi(111) on Si(111) after deposition of 0.8 BL Sb (**b**) and 1.5 BL Sb (**c**).

From topographic images obtained through repeated STM experiments, we have deduced the relative areas occupied by Bi clean surface, 1BL antimonene, and 2BL antimonene for different nominal coverages θ of Sb (Fig. C2). The relative area of 1BL antimonene amounts to ~70% of the whole surface for θ = 1.0 BL, then decreases to ~50% for θ = 1.5 BL. By contrast, the relative area of 2BL antimonene increases from ~20% to ~50% as nominal coverage increases from θ = 1.0 to 1.5 BL. By extrapolating these trends to θ = 2.0 BL, the relative areas of 1BL antimonene can be estimated to be 20% and that of 2BL antimonene 70%. Therefore, we can safely conclude that 1BL and 2BL antimonenes are the dominant phases for θ = 1.0 BL and 2.0 BL, respectively. Since the ARPES signal intensity is proportional to the surface area to be detected, the band structures obtained for these nominal coverages mostly reflect those of pure 1BL and 2BL antimonenes.

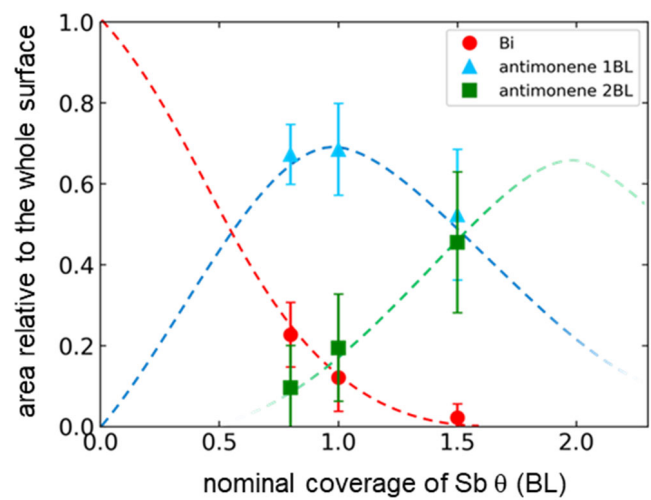

**Figure C2 | Relative areas of antimonene grown on a Bi(111) surface.** The areas of Bi clean surface, 1BL antimonene, and 2BL antimonene relative to the whole surface are displayed as a function of nominal coverage θ of Sb.  Dashed lines are eye guides.

## D. ARPES/SARPES measurements of antimonene on a Bi(111) surface

We performed ARPES/SARPES measurements of 2BL Sb on a Bi(111) surface. The results are nearly identical to those of 1BL Sb/Bi(111) (Fig. 3b, d, e); the $S_1$ and $S_2$ bands dispersing from the $\bar{\Gamma}$ point similarly (Fig. D1a), Rashba-type spin polarisation (Fig. D1b), and saddle points on the Fermi surface along the $\bar{\Gamma} - \bar{K}$ direction (the red ellipsoids in Fig. D1c). The van Hove singularity of the saddle points also explains the $dI/dV$ spectral peaks observed for 2BL Sb/Bi(111) (Fig. 2e).

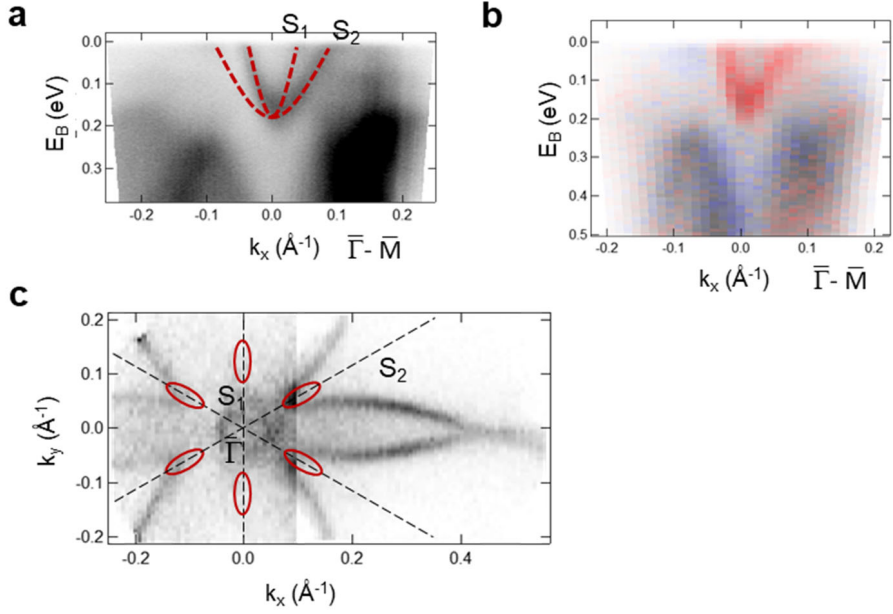

**Fig. D1 | Electronic and spin structures of 2BL antimonene/Bi(111) moiré superlattices in the momentum space. a** 2D plot of ARPES intensity as a function of momentum $k_x$ along the $\bar{\Gamma} - \bar{M}$ direction and as a binding energy $E_B$. **b** 2D plot of the SARPES signal as a function of momentum $k_x$ along the $\bar{\Gamma} - \bar{M}$ direction and as a binding energy $E_B$. The intensity and the spin polarization in the y direction are indicated by brightness (dark: high, bright: low) and colour (red: positive, blue: negative), respectively. **c** 2D plot of ARPES intensity measured near the Femi level ($E_B$ = 0.03 eV) in the $k_x$ - $k_y$ space, which gives the Fermi surface contour. The plot includes two data sets obtained in different runs, which causes an apparent discontinuity at $k_x$ = 0.12 Å⁻¹. The dashed lines indicate the $\bar{\Gamma} - \bar{K}$ directions. The red ellipsoids show the locations of the saddle points of $S_2$ band.

In Figs. D2a and D2b, we redisplay the same ARPES data of Figs. 3b and 3c without the eye guides to have a better view of the band structure. Although the photoemission intensity is very weak in some areas, we can safely restore the total band structure as in Fig. 3 by considering the symmetry with respect to the Γ point (due to the $C_3$ and time-reversal symmetries of the present system).

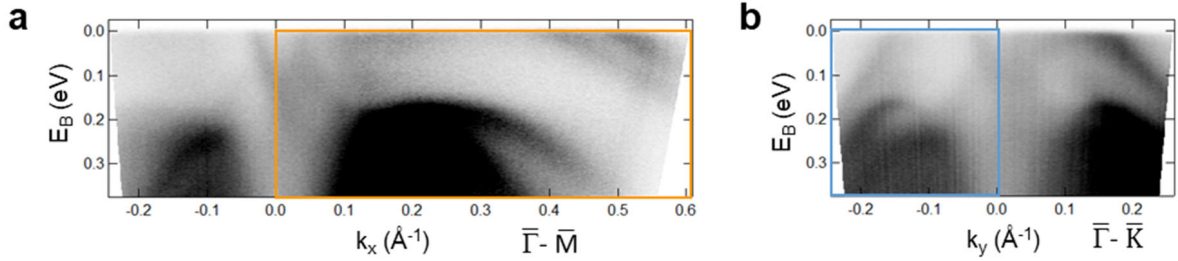

**Fig. D2 | Electronic structures of 1BL antimonene/Bi(111) moiré superlattices in the momentum space (without eye guide). a** 2D plot of ARPES intensity as a function of momentum $k_x$ along the $\bar{\Gamma} - \bar{M}$ direction and as a binding energy $E_B$. **b** The same plot as in **b** along the $\bar{\Gamma} - \bar{K}$ direction. The orange and blue rectangles correspond to those shown in Fig. 4a. The data are identical to those of Figs. 3b and 3c.

In Fig. D3, we redisplay the same SARPES data of Fig. 3g without the eye guides for the Fermi surface contour. The ellipses A, B, C, D and A', B', C', D' show the areas near the $\bar{\Gamma} - \bar{K}$ lines at ±60° to the $k_x$ axis (black dashed lines), where spin polarization is reversed when momentum is reversed with respect to the $\bar{\Gamma}$ point. The variations in signal intensity can be attributed to the fact that SARPES/ARPES signal intensity is strongly dependent on the transition matrix in the photoemission process.

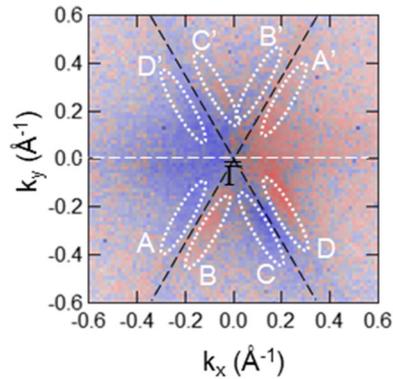

**Fig. D3 | Spin structures of 1BL antimonene/Bi(111) moiré superlattices in the momentum space.** The figure shows a 2D plot of SARPES intensity and spin polarization in the y direction measured near the Fermi level ($E_B$ = 0.03 eV). The data are identical to those of Fig. 3g.

## E. DFT band structure calculations with epitaxial Sb(111)/Bi(111) models

First, moiré superlattices of 1BL Sb(111)/5BL and 2BL Sb(111)/5BL were numerically optimized to determine the locations of the Sb atoms in the out-of-plane direction at the centers of the AA, AB and AC stacking regions (indicated by the red, blue and green circles, respectively; see Fig. E1). These structural parameters were used to construct the epitaxial models of 1BL Sb(111)/5BL Bi(111) and 2BL Sb(111)/5BL Bi(111), each with AA, AB and AC stackings.

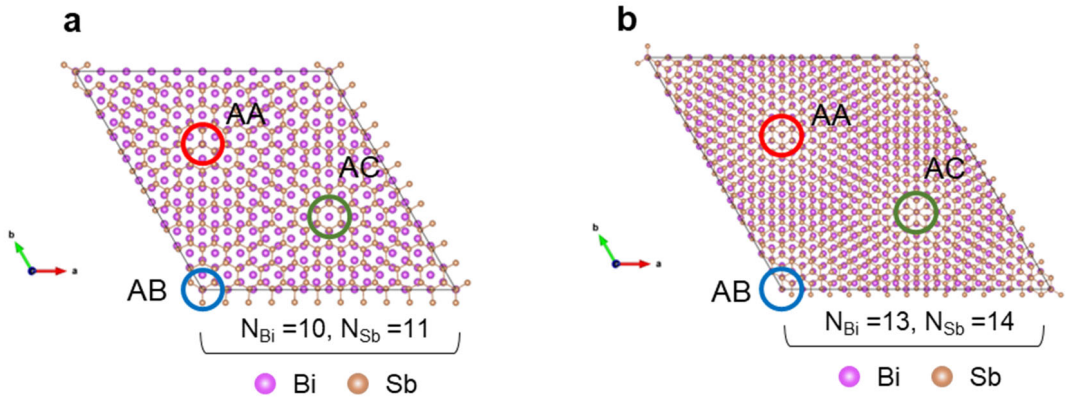

**Fig. E1 | Atomic structure model for moiré superlattices. a** Moiré superlattice of 1BL Sb(111)/5BL, which consists of a single 11×11 supercell of the Sb(111) BL on five vertically stacked 10×10 supercells of the Bi(111) BL. **b** Moiré superlattice of 2BL Sb(111)/5BL, which consists of a double 14×14 supercell of the Sb(111) BL on five vertically stacked 13×13 supercells of the Bi(111) BL.

The DFT calculations were conducted for an epitaxial model of 1BL Sb(111) on 5BL Bi(111) and for an epitaxial model of 2BL Sb(111) on 5BL Bi(111). Here the results for the AA (Fig. E2) and AC (Fig. E3) stackings are displayed. Regarding the results for the AB stacking, see Fig. 4 in the main text. The energies and momenta of the saddle points of the $S_2$ band are shown in Table E1.

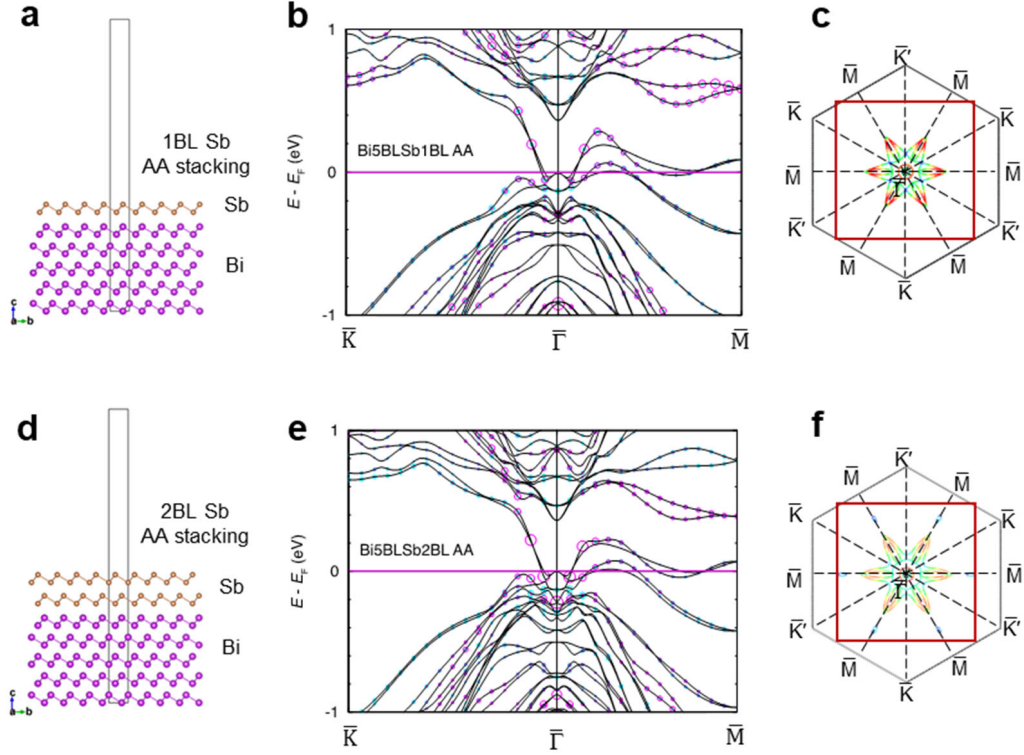

**Fig. E2 | DFT band structure calculations for the Sb(111)/Bi(111) epitaxial models with the AA stacking. a, d** Atomic structure models for 1BL Sb(111)/5BL Bi(111) (**a**) and 2BL Sb(111)/5BL Bi(111) (**d**). The solid rectangles are the unit cells for calculations. **b, e** Band dispersions along the $\overline{K} - \overline{\Gamma} - \overline{M}$ direction calculated for 1BL Sb(111)/5BL Bi(111) (**b**) and 2BL Sb(111)/5BL Bi(111) (**e**). **c, f** Fermi surfaces calculated for 1BL Sb(111)/5BL Bi(111) (**c**) and 2BL Sb(111)/5BL Bi(111) (**f**). The AA stacking sequence was taken for all models.

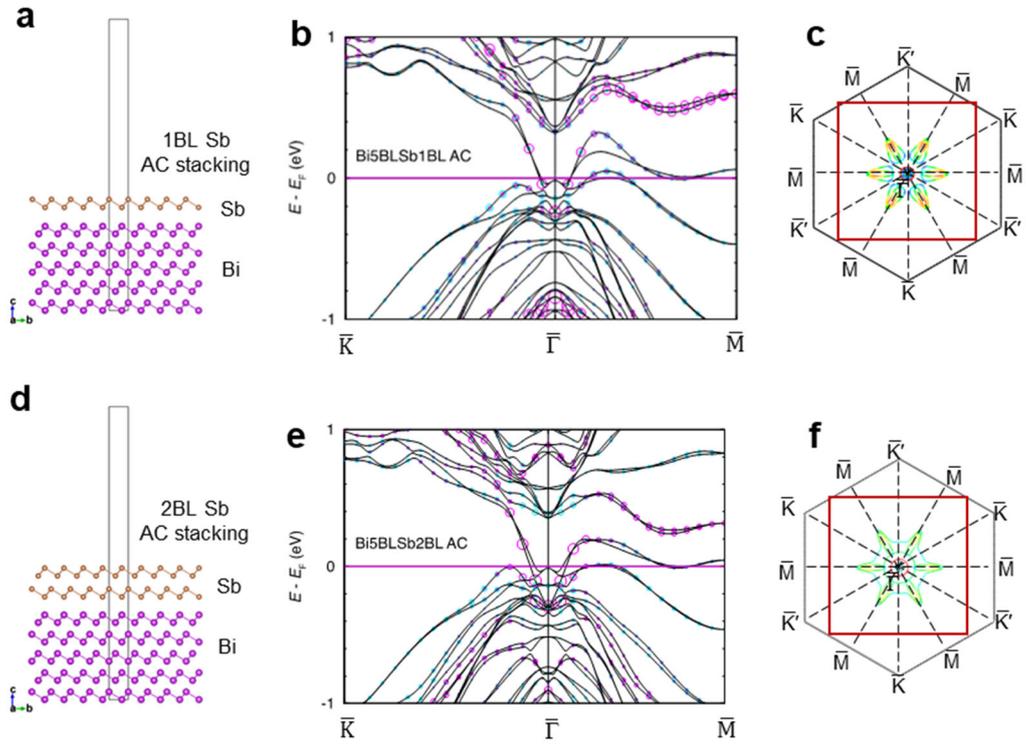

**Fig. E3 │ DFT band structure calculations for the Sb(111)/Bi(111) epitaxial models with the AC stacking. a, d** Atomic structure models for 1BL Sb(111)/5BL Bi(111) (**a**) and 2BL Sb(111)/5BL Bi(111) (**d**). The solid rectangles are the unit cells for calculations. **b, e** Band dispersions along the $\overline{K} - \overline{\Gamma} - \overline{M}$ direction calculated for 1BL Sb(111)/5BL Bi(111) (**b**) and 2BL Sb(111)/5BL Bi(111) (**e**). **c, f** Fermi surfaces calculated for 1BL Sb(111)/5BL Bi(111) (**c**) and 2BL Sb(111)/5BL Bi(111) (**f**). The AC stacking sequence was taken for all models.

| model and stacking | 1BL Sb(111)/5BL Bi(111) | | |
|---|---|---|---|
| | AA | AB | AC |
| $E$-$E_F$ (meV) | -41 | -57 | -48 |
| $(k_x, k_y)$ (Å$^{-1}$) | (0, 0.155) | (0, 0.173) | (0, 0.155) |
| model and stacking | 2BL Sb(111)/5BL Bi(111) | | |
| | AA | AB | AC |
| $E$-$E_F$ (meV) | -30 | -32 | -3 |
| $(k_x, k_y)$ (Å$^{-1}$) | (0, 0.144) | (0, 0.144) | (-0.035, 0.166) (0.035, 0.166) |

**Table E1 | The energies ($E$-$E_F$) and momenta ($k_x$, $k_y$) of the saddle points of the S$_2$ band.** Numerical errors in energy and momentum are estimated to be ±2 meV and ± 0.007 Å$^{-1}$, respectively. Here, the $k_x$ ($k_y$) axes are set along the $\overline{\Gamma} - \overline{M}$ ($\overline{\Gamma} - \overline{K}$) direction (see Fig. 3a). All the saddle points are located along the $\overline{\Gamma} - \overline{K}$ lines except for the split pair for 2BL Sb(111)/5BL Bi(111) with the AC stacking. Note that there exist additional saddle points due to the six-fold symmetry of the Brillouin zone.

Figure E4a shows the spin structure at the Fermi surface of the 1BL Sb(111)/5BL Bi(111) epitaxial model with the AB stacking. The spin polarizations at both $S_1$ and $S_2$ bands are mostly aligned in the tangential directions on the Fermi contour and have opposite vorticities, in line with the Rashba-type spin splitting. As expected, the spin polarization in the y direction $P_y$ is reversed with respect to the $\bar{\Gamma}$ point. However, since the Fermi surface of the $S_2$ band is strongly deformed to a star-like shape, $P_y$ is reversed with respect to the the $\bar{\Gamma} - \bar{K}$ lines at $\pm 60°$ to the $k_x$ axis (dashed lines). They are consistent with the SARPES result in Fig. 3g (see also Supplementary Information D, Fig. D3). The same features are also confirmed for the spin structures of the 2BL Sb(111)/5BL Bi(111) model (Fig. E4b).

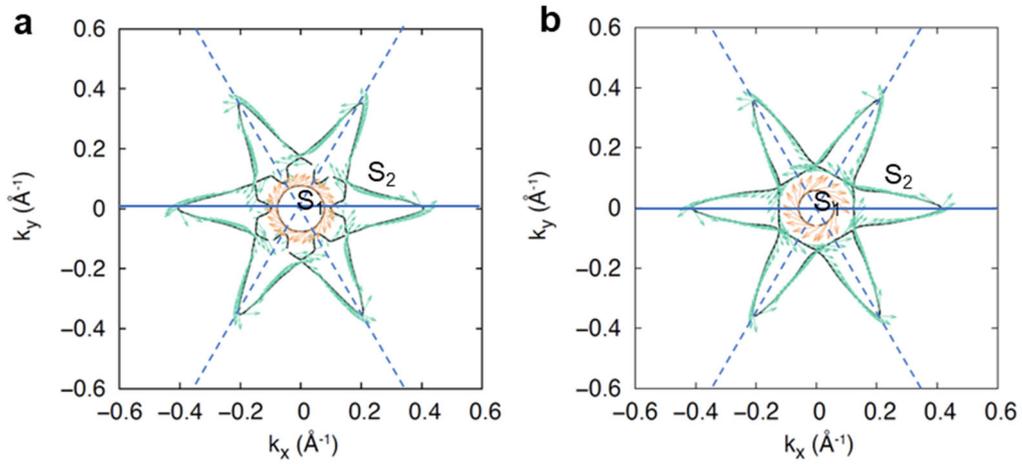

**Fig. E4 | Spin structures at the Fermi surface of Sb(111)/Bi(111) epitaxial model. a,** Spin polarization directions on the $S_1$ band (orange) and the S2 band (green) calculated for the 1BL Sb(111)/5BL Bi(111) epitaxial model with the AB stacking. The solid and dashed lines indicated the $\bar{\Gamma} - \bar{K}$ lines at 0° and ±60° to the $k_x$ axis, respectively. **b,** The result of the same calculations for the 2BL Sb(111)/5BL Bi(111) epitaxial model with the AB stacking

## F. DFT band structure calculations with a moiré superlattice model

To investigate the influence of the moiré superlattice on the electronic band structure, we conducted DFT calculations based on a moiré model of 1BL Sb(111)/3BL Bi(111), which consists of a single 11×11 supercell of the Sb(111) BL on three vertically stacked 10×10 supercells of the Bi(111) BL. Here, the number of Bi(111) bilayers was reduced from that of the epitaxial model to enable the full calculations. The moiré superlattice was optimized in the same way as described in the main text (see Methods). Figure F1a shows the calculated band structure along the $\overline{\Gamma} - \overline{K}_{moiré}$ line, where $\overline{K}_{moiré}$ belongs to the moiré Brillouin. The figure indicates the absence of moiré-induced energy gap at the zone boundary. As a reference, the band structure along the $\overline{\Gamma} - \overline{K}$ line of an epitaxial model of 1BL Sb(111)/3Bi(111) is shown in Fig. F1b.

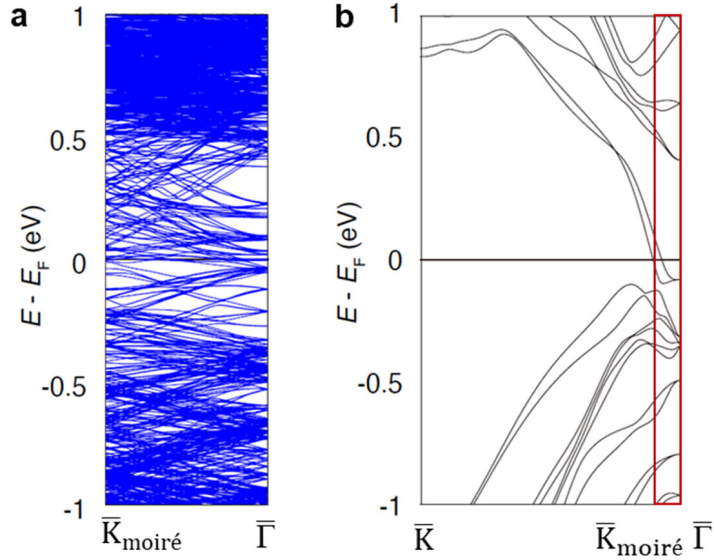

**Fig. F1 | Electronic band structure of Sb(111)/Bi(111) moiré model. a** The calculated band structure of a moiré model of 1BL Sb(111)/3BL Bi(111), which consists of a single 11×11 supercell of the Sb(111) BL on three vertically stacked 10×10 supercells of the Bi(111) BL. The figure shows the band dispersion along the $\overline{\Gamma} - \overline{K}_{moiré}$ line, where $\overline{K}_{moiré}$ belongs to the moiré Brillouin. **b** The band structure along the $\overline{\Gamma} - \overline{K}$ line of an epitaxial model of 1BL Sb(111)/3Bi(111). The red square corresponds to the region displayed in **a**.